\documentclass[conference]{IEEEtran}

\IEEEoverridecommandlockouts

\usepackage{cite}
\usepackage{amsmath,amssymb,amsfonts}
\usepackage{algorithmic}
\usepackage{graphicx}
\usepackage{textcomp}
\usepackage{xcolor}
\usepackage{booktabs}
\def\BibTeX{{\rm B\kern-.05em{\sc i\kern-.025em b}\kern-.08em
    T\kern-.1667em\lower.7ex\hbox{E}\kern-.125emX}}
\begin{document}

\title{All I Hear is Noise? Investigating Clever Hans Effects in Clinical Speech Datasets
}

\author{\IEEEauthorblockN{Melanie Jouaiti}
\IEEEauthorblockA{\textit{School of Computer Science} \\
\textit{University of Birmingham}\\
Birmingham, UK \\
m.jouaiti@bham.ac.uk}
\and
\IEEEauthorblockN{Ning Ma}
\IEEEauthorblockA{\textit{School of Computer Science} \\
\textit{University of Sheffield}\\
Sheffield, UK \\
n.ma@sheffield.ac.uk}
}

\maketitle

\begin{abstract} 
Recent work revealed a striking Clever Hans effect in the Pitt dataset, where Alzheimer’s detection achieved nearly 100\% accuracy using only silent audio segments. This raises serious concerns about hidden confounding factors in speech-based health datasets. We systematically investigate whether similar biases exist across five widely used clinical speech corpora: DAIC-WoZ (depression), TORGO (dysarthria), Neurovoz (PD), MDVR-KCL (PD), and UCLASS (stuttering). For each dataset, we compare classification using the first second of audio, silent segments, and full recordings, and evaluate both raw and denoised signals. Across all datasets, silence-only classification frequently matched or exceeded full-audio performance, suggesting that classification performance may be influenced by dataset-specific confounds in addition to disorder-related speech characteristics. These findings question the reliability and generalisability of speech-based biomarkers and call for stricter methodological reporting, preprocessing transparency, and bias mitigation.
\end{abstract}


\section{Introduction}
\label{sec:intro}

Speech is increasingly proposed as a non-invasive, scalable digital biomarker for neurological and psychiatric conditions, including depression, Parkinson’s disease, dementia, and dysarthria~\cite{cummins2015review,voleti2019review,robin2020evaluation,de2020artificial,bowden2023systematic}. Advances in machine learning and speech representation learning have demonstrated promising performance in detecting clinical conditions from speech recordings. These approaches typically rely either on engineered acoustic and linguistic features classified using machine learning models, or on end-to-end deep learning models trained directly on speech signals. While these paradigms differ in interpretability and modelling assumptions, both aim to capture disorder-related characteristics embedded in speech.

However, clinical speech datasets are often collected under heterogeneous, often unstandardised recording conditions, using different microphones, recording environments, preprocessing pipelines, and dataset-specific collection protocols. Such variability can introduce unintended correlations between diagnostic labels and non-clinical factors. As a result, models may achieve strong performance by exploiting spurious cues that are correlated with the target labels rather than learning clinically meaningful speech characteristics.
This phenomenon is closely related to shortcut learning and the Clever Hans effect, where models rely on unintended but predictive signals instead of the intended underlying phenomenon~\cite{geirhos2020shortcut,torralba2011unbiased}. In healthcare applications, such behaviour is particularly concerning
as models that exploit dataset-specific artefacts may fail to generalise beyond the conditions under which the data were collected~\cite{goetz2024generalization}.

Recent evidence has highlighted this risk in speech-based health research. Liu et al.~\cite{liu2024clever} reported the Clever Hans effect in the Pitt corpus for Alzheimer's disease detection, where classification achieved nearly 100\% accuracy using only silent portions of the recordings. Similarly, Schu et al.~\cite{schu2023dysarthric} demonstrated that state-of-the-art dysarthria classification models using the UA-Speech and TORGO databases often learn environmental recording artifacts rather than disorder-specific speech characteristics. 
Despite growing awareness of shortcut learning in speech-based clinical research, it remains unclear whether such effects are limited to a single dataset or represent a broader challenge across clinical speech datasets. Many widely used datasets continue to serve as benchmarks for developing and evaluating speech-based diagnostic systems, yet systematic investigation of potential Clever Hans effects remains limited. Establishing the prevalence of such effects is therefore essential for assessing the reliability and generalisability of speech-based biomarkers.

In this work, we examine whether predictive information unrelated to the intended speech signal is present in five widely used clinical speech datasets: DAIC-WoZ (depression), TORGO (dysarthria), Neurovoz (Parkinson’s disease), MDVR-KCL (Parkinson’s disease), and UCLASS (stuttering). 
We evaluate model performance under deliberately constrained conditions using (i) only the first second of each recording, (ii) non-speech segments identified by voice activity detection, and (iii) the full audio recording. We also compare performance when using the raw and the denoised audio to assess the impact of recording artefacts. By extending Clever Hans analysis beyond a single dataset and disorder, we aim to determine 
the extent to which unintended dataset artefacts may contribute to reported classification performance and to highlight the need for more rigorous evaluation and reporting practices in speech-based healthcare research. Throughout this work, strong classification performance obtained from non-speech segments or very short excerpts is interpreted as evidence that models may be exploiting information beyond the intended disorder-related speech characteristics.

\section{Related work}

\subsection{Clever Hans effect}

The Clever Hans effect refers to situations in which machine learning models achieve strong performance by exploiting unintended or task-irrelevant features that correlate with target labels in the dataset rather than learning the intended underlying phenomenon~\cite{geirhos2020shortcut}. Closely related to shortcut learning and dataset bias, this issue has been widely documented across domains.
It is particularly concerning in health contexts, where any model needs to be rigorously validated to ensure reliability, generalisability, and effectiveness beyond the specific conditions under which data were collected.

For example, in medical imaging, Haynes et al.~\cite{haynes2024generalisation} showed that COVID-19 classification performance was not affected by occluding the lung regions in Chest X-Ray images for certain models pre-trained on the ImageNet dataset and fine-tuned on COVID-19 data, which suggests reliance on non-pathological artefacts. In speech-based dementia detection, Liu et al.~\cite{liu2024clever} reported that Alzheimer’s disease classification on the Pitt corpus achieved nearly 100\% accuracy using only silent segments of recordings. This striking result indicates that non-speech cues, such as recording conditions or channel characteristics, can encode strong predictive signal independent of speech production.

Various strategies have been proposed to mitigate shortcut learning, including data augmentation and domain randomisation to reduce spurious correlations~\cite{shorten2019survey}, adversarial training to discourage reliance on confounds~\cite{zhang2018mitigating}, and interpretability techniques such as attention visualisation to better understand model behaviour~\cite{ribeiro2016model}. Models can also be forced to concentrate on more relevant features by utilising attention maps~\cite{halawani2025enhanced,wu2022classification}. However, systematic auditing of clinical speech datasets for Clever Hans effects remains limited.

\subsection{Depression detection}

The prevalence of depression is on the rise, making it a global public health concern \cite{lepine2011increasing}. 
According to estimates from the World Health Organization (2023), depression is one of the most common illnesses in the world, affecting around 280 million people and accounting for 3.8\% of the global population. Depression is characterised by both purely psychological and emotional symptoms, such as low mood, diminished interest in all activities, feelings of worthlessness, difficulty concentrating, and recurrent thoughts of death, as well as physical symptoms, such as changes in body weight, sleep patterns, and psychomotor changes~\cite{american2015american}. 
Beyond affective symptoms, depression is associated with measurable changes in speech and language, including alterations in prosody, lexical choice, syntactic complexity, and discourse structure \cite{cummins2015review}.

Automatic depression detection has therefore become a prominent application of speech and language processing, particularly using benchmark datasets such as DAIC-WoZ~\cite{gratch2014distress}.
Studies have successfully employed machine learning techniques to discriminate healthy subjects from subjects with depression by analysing speech transcripts~\cite{rohanian2019detecting, ilias2024cross, li2023using}. Language and textual analysis, taking into account linguistic attributes, such as grammar, syntax, and the vocabulary of depressed subjects, have also proven to be accurate, dependable, and objective methods capable of differentiating depressed people from healthy participants~\cite{smirnova2018language, esposito2020biological}.

While reported results are promising, most studies focus primarily on predictive accuracy. Comparatively less attention has been paid to the potential influence of recording conditions, interviewer effects, or other dataset-specific artefacts that may confound model performance.

\subsection{Dysarthria detection}

Dysarthria is caused by a neurological disorder that affect muscle control and, in turn, breathing, vocalisation, resonance, articulation, and prosody \cite{enderby2013disorders}. Volume tends to be inappropriate due to damage to the central and peripheral nervous system. Dysarthria is present in neurological conditions, such as stroke, Parkinson’s disease, brain trauma, brain tumours, cerebral palsy, amyotrophic lateral sclerosis, multiple sclerosis, muscular dystrophy. Hoarseness and monotonicity are also common \cite{darley1969differential}.
Timely intervention and early rehabilitation are critical for course management, and to avoid associated social difficulties, depression, and other psychological problems~\cite{brady2011impact}

Early studies explored handcrafted features such as pitch period entropy~\cite{vashkevich2018features}, glottal to noise excitation and formant frequency~\cite{muhammad2017voice}, Mel-frequency cepstral coefficients (MFCC)~\cite{jothieswari2024enhancing}, and perception linear predictive coefficients~\cite{polur2005experiments}. More recently, deep learning approaches, such as convolutional neural network (CNN), CNN-LSTM (long short-term memory), have also been successfully employed~\cite{sajiha2024automatic, hassan2025enhanced}. Despite promising results, few studies explicitly evaluate whether classification performance may be influenced by recording conditions or other confounding factors.

\subsection{Parkinson's detection}

Parkinson’s Disease (PD) is a progressive neurodegenerative disorder, primarily associated with motor impairments affecting muscle control, balance and movement~\cite{bloem2021parkinson}. Motor symptoms also have an impact on communication, as muscle rigidity can affect speech and facial expressions. 
Speech changes commonly include impairments in loudness of voice, pitch, articulation, speech rate and prosody~\cite{schalling2018speech}. Hypokinetic dysarthria can also be present. 
In addition, cognitive slowing may affect language processing and conversational fluency, contributing to word-finding difficulties and reduced communicative efficiency. Speech impairment is among the top four concerns of people with Parkinson's disease~\cite{schalling2018speech}.

Many different approaches have been investigated for PD classification on the NeuroVoz dataset: using burst segments (82\% accuracy) \cite{moro2017use}, using speaker recognition and allophonic grouping (86\% accuracy) \cite{moro2018study}, using  phonatory features (89\% accuracy) \cite{moro2019analysis}, articulation analysis (81\%-89\%) \cite{moro2019forced, moro2019phonetic}. Godino et al. \cite{godino2019approaches} used various machine learning approaches, reporting accuracies over 85\%. While these findings demonstrate discriminative potential, systematic evaluation of shortcut learning or dataset bias in these corpora remains scarce.

\section{Methods}
\label{sec:typestyle}

\subsection{Datasets}

\begin{table*}[t] 
\caption{Summary of the clinical speech datasets investigated in this study.} \label{tab:datasets} 
\centering 
\vspace{-.2cm}
\begin{tabular}{lllrll} 
\toprule 
Dataset & Disorder & Task & Speakers & Language & Recording Type \\ 
\midrule 
DAIC-WOZ & Depression & Depressed / Non-depressed & 275 & English & Clinical interview \\ 
TORGO & Dysarthria & Dysarthric / Control & 15 & English & Read and prompted speech \\ 
Neurovoz & Parkinson's disease & PD / Control & 112 & Spanish & Speech tasks and monologue \\ 
MDVR-KCL & Parkinson's disease & PD / Control & 37 & English & Reading and dialogue \\ 
UCLASS & Stuttering & Recording environment$^\dagger$ & 128 & English & Spontaneous speech \\ 
\bottomrule
\multicolumn{6}{p{15cm}}{\footnotesize $^\dagger$UCLASS contains only speakers who stutter and does not include healthy controls. We therefore classify recording environment rather than clinical status.}
\end{tabular} 
\end{table*}

We evaluate five widely used clinical speech datasets spanning different disorders and recording conditions. \textbf{Extended DAIC} \cite{gratch2014distress} contains audio recordings and transcripts of 275 English clinical interviews conducted with a human and with a virtual agent. Each participant is annotated with a PHQ-8 score and a binary depression label (PHQ-8 $\ge$ 10 indicates depression). The dataset is split into training (163 participants; 37 depressed, 126 non-depressed), development (56; 12 depressed, 44 non-depressed), and test sets (56; 17 depressed, 39 non-depressed). Automatic depression detection is typically evaluated on two tasks: predicting depression severity and classification of depression presence. 

\textbf{TORGO} \cite{rudzicz2012torgo} contains audio recordings by the University of Toronto from seven speakers with dysarthria and eight speakers without dysarthria. Speakers with dysarthria have cerebral palsy or amyotrophic lateral sclerosis. The dataset includes read and prompted speech under controlled recording conditions.

\textbf{Neurovoz} \cite{mendes2024neurovoz} contains audio recordings from 112 native Castilian-Spanish speakers, including 58 healthy controls and 54 individuals with PD. Tasks include sustained vowels, diadochokinetic tests, 16 Listen-and-Repeat utterances, and spontaneous monologues. The protocol reports that all participants were recorded under the same environmental conditions and using the same equipment. 
This dataset has achieved a benchmark accuracy of 89\% for the screening of PD \cite{mendes2024neurovoz}. 

\textbf{MDVR-KCL} \cite{jaeger2019mobile} contains audio recordings from 21 healthy controls and 16 PD patients. Each of the 37 participants performed two tasks: reading a paragraph of text and engaging in spontaneous dialogue. Data was recorded with a smartphone.

\textbf{UCLASS} \cite{howell2009university} contains recordings from 128 children and adults who stutter. This dataset does not contain healthy speech. It does, however, provide with recording setting information, such as recorded in the clinic or at University College London. For this dataset, we attempt to classify the recording setting to examine whether environmental factors are predictable from the audio signal.

\subsection{Features}

All audio recordings are resampled to 16\,kHz prior to feature extraction. Two complementary feature representations are extracted: \textbf{OpenSMILE} \cite{eyben2010opensmile} is a feature extraction tool relevant for machine learning applications. It has been extensively used in clinical speech analysis \cite{lombardo2025speech}, as well as paralinguistic challenges. The extended Geneva Minimalistic Acoustic Parameter Set (eGeMAPSv02) configuration \cite{xue2019acoustic} is a comprehensive set of 88 voice parameters including frequency, amplitude, spectral, cepstral, and temporal features. The Compare 16 configuration is even more extensive with 6373 parameters. Despite the concerns about its robustness and reliability \cite{choi2025comparative}, it has become a standard in clinical speech research.

\textbf{wav2vec 2.0} is a transformer-based speech model that has shown success in phonetics-related tasks, including phoneme recognition, tone recognition, and speaker identification. It encodes various types of information, with different layers capturing different aspects. 
Wav2vec 2.0 embeddings have been used for the classification of pathological speech \cite{cai2024voice, javanmardi2023wav2vec}.

\subsection{Processing}

To investigate whether predictive information is present outside the intended speech signal, we evaluated three input conditions: (i) the first second of each recording, (ii) the full recording, and (iii) non-speech segments only. Non-speech segments were identified using the Silero Voice Activity Detection (VAD) model (initial parameters: threshold=0.8,
          window\_size\_samples=128,
          min\_silence\_duration\_ms = 50,
          return\_seconds=False,
          speech\_pad\_ms = 30; if no speech is detected, threshold is iteratively decreased until speech is detected). All regions classified as non-speech were concatenated prior to feature extraction. This procedure preserves recording-level acoustic characteristics while removing voiced speech content. Feature extraction was then performed on the resulting concatenated non-speech signal using the same pipeline as for the full recordings.


To assess the prevalence of non-relevant speech information  at the beginning of recordings, we quantified the proportion of samples for which the first second contained no detected speech using VAD or aligned transcripts if available. The percentages were 85\% for TORGO, 84\% for DAIC-WOZ (in the remaining 16\%, the voice of the experimenter can often be heard in the first second), 26\% for UCLASS, 65\% for MDVR-KCL reading (manually checking the files reveals that the first second is either silence, setup noises or faint experiment speech for all the files), 61\% for MDVR-KCL dialogue (manually checking the files reveals that the first second is either silence or faint experiment speech for all the files), and 10\% for Neurovoz. These results indicate that the first second frequently contains little or no relevant participant speech content in most datasets, which provides a useful test case for assessing the contribution of non-speech information to classification performance.

We additionally evaluated all conditions using both raw and denoised audio. Denoising was performed using the \textit{noisereduce} Python library, which applies spectral-gating-based noise suppression. Comparing raw and denoised signals allows us to assess the extent to which recording artefacts and background noise contribute to predictive performance.

\begin{table*}[ht!]
    \centering
    \caption{Weighted F1-scores using OpenSMILE Compare16 features across datasets. Baseline results from prior literature are included where available. Bold values indicate the highest score per dataset. UCLASS results correspond to recording-site classification rather than clinical classification.}
    \vspace{-.2cm}
    \begin{tabular}{c|c|c|c|c|c|c|c}
    \toprule
          & 1st sec & full & silence & denoised 1st sec & denoised full & denoised silence & baseline \\
    \midrule
    TORGO & 0.87 & 0.72 & 0.82 & \textbf{0.91} & 0.90 & 0.74 & 0.98 \cite{remya2025hybrid}\\
    DAIC  & 0.66 & 0.66 & 0.67 & 0.61 & 0.61 & \textbf{0.72} & 0.80 \cite{ravi2024enhancing}\\
   UCLASS & 0.60 & 0.76 & \textbf{0.83} & 0.67 & 0.75 & 0.61 & N/A \\
   MDVR-KCL & 0.67 & \textbf{0.87} & 0.52 & 0.54 & 0.73 & \textbf{0.87} & 0.92 \cite{di2024machine}\\
 NeuroVoz & \textbf{0.72} & \textbf{0.72} & 0.71 & 0.65 & 0.68 & 0.54  & 0.72 \cite{gomez2025discriminating}\\
 \bottomrule
    \end{tabular}
    \label{tab:f1-compare16}
\end{table*}

\begin{table*}[ht!]
    \centering
    \caption{Weighted F1-scores using OpenSMILE eGeMAPSv02 features (88 acoustic descriptors). For reminder, for UCLASS, we classify where the audio was recorded}
    \vspace{-.2cm}
    \begin{tabular}{c|c|c|c|c|c|c|c}
    \toprule
          & 1st sec & full & silence & denoised 1st sec & denoised full & denoised silence & baseline \\ 
    \midrule
    TORGO & 0.82 & 0.82 & \textbf{0.90} & 0.85 & 0.84 & 0.74 & 0.98 \cite{remya2025hybrid} \\
    DAIC  & \textbf{0.75} & \textbf{0.75} & 0.66 & 0.62 & 0.62 & 0.60 & 0.80 \cite{ravi2024enhancing} \\
   UCLASS & 0.61 & 0.68 & 0.71 & 0.66 & \textbf{0.84} & 0.66 & N/A \\
    MDVR-KCL & 0.67 & 0.74 & 0.59 & 0.60 & 0.74 & 0.60 & 0.92 \cite{di2024machine} \\
 NeuroVoz & 0.68 & 0.69 & 0.68 & 0.68 & \textbf{0.71} & 0.48 & 0.72 \cite{gomez2025discriminating} \\
 \bottomrule
    \end{tabular}
    \label{tab:f1-eGeMAPSv02}
\end{table*}

\begin{table*}[ht!]
    \centering
    \caption{Weighted F1-scores using wav2vec 2.0 embeddings.}
    \vspace{-.2cm}
    \begin{tabular}{c|c|c|c|c|c|c}
    \toprule
          & 1st sec & full & silence & denoised 1st sec & denoised full & denoised silence \\
    \midrule
    TORGO & 0.75 & 0.87 & 0.90 & \textbf{0.91} & \textbf{0.91} & 0.71 \\
    DAIC  & \textbf{0.74} & \textbf{0.74} & \textbf{0.74} & 0.68 & 0.69 & 0.69 \\
   UCLASS & 0.71 & 0.85 & \textbf{0.89} & 0.73 & 0.85 & 0.66 \\
 MDVR-KCL & 0.54 & \textbf{0.74} & 0.46 & 0.49 & \textbf{0.74} & 0.54 \\
 NeuroVoz & 0.72 & \textbf{0.75} & 0.65 & 0.68 & 0.74 & 0.62  \\
    \bottomrule
    \end{tabular}
    \label{tab:f1-wav2vec}
\end{table*}

\subsection{Classification}

Classification is performed using a Gradient Boosting classifier with 100 estimators, a learning rate of 1.0, a maximum depth of 5, random state 0, and balanced sample weights. Our objective is not to optimise classification accuracy, but rather to assess whether predictive information remains available under increasingly constrained input conditions. Consequently, the same classifier and evaluation protocol are applied throughout all experiments.
To minimise speaker leakage, all experiments employed consistent speaker-independent splits, with
approximately 80\% of speakers used for training and 20\% for testing. Although standard evaluation protocols differ across datasets (e.g., leave-one-subject-out evaluation for TORGO and predefined training/test partitions for DAIC-WOZ), a common speaker-independent evaluation protocol was adopted for all datasets to facilitate a controlled comparison of Clever Hans effects.
Performance is reported using a weighted average F1-score.

\section{Results and Discussion} \label{sec:results}

Weighted F1-score results using different feature representations (OpenSMILE Compare16, OpenSMILE eGeMAPS, and wav2vec 2.0 embeddings) across all five datasets are reported in Tables~\ref{tab:f1-compare16}-\ref{tab:f1-wav2vec}. Across all three feature representations, classification based on the first second of audio frequently matches or exceeds performance obtained using the full recording. For several datasets, performance based solely on silent or unvoiced segments remains high and in some cases surpasses full-audio classification. For example, UCLASS achieves F1 = 0.89 (Wav2Vec) when using silence only. Similarly, wav2vec embeddings exhibit strong performance when trained on non-speech portions.

If disorder-specific speech characteristics were the primary discriminative signal, one would expect performance to improve with access to longer recordings and degrade substantially when limited to unvoiced or silent segments. Instead, full-audio classification does not consistently outperform first-second or unvoiced-only conditions across datasets. Given that the first second frequently contains silence, experimenter speech, or setup noise in several datasets, it is unlikely to provide sufficient disorder-related speech information for reliable classification.
These findings suggest that models may be exploiting information beyond the intended disorder-related speech signal, potentially including background noise, channel characteristics, recording artefacts, or other dataset-specific confounds. 

Denoising sometimes improves performance, particularly when classifying short or silent segments (e.g., DAIC, MDVR-KCL). For TORGO, the highest F1-score is achieved using denoised first-second inputs. However, denoising does not uniformly improve results; for example, performance drops for silent segments in Neurovoz. This variability suggests that models may rely on subtle recording characteristics that are altered or removed during preprocessing.

For UCLASS, we are also able to infer the recording site with an F1 score up to 0.89, confirming that extracted features are strongly corrupted by site characteristics.

\begin{figure*}[ht!]
    \centering
    \includegraphics[width=0.32\linewidth]{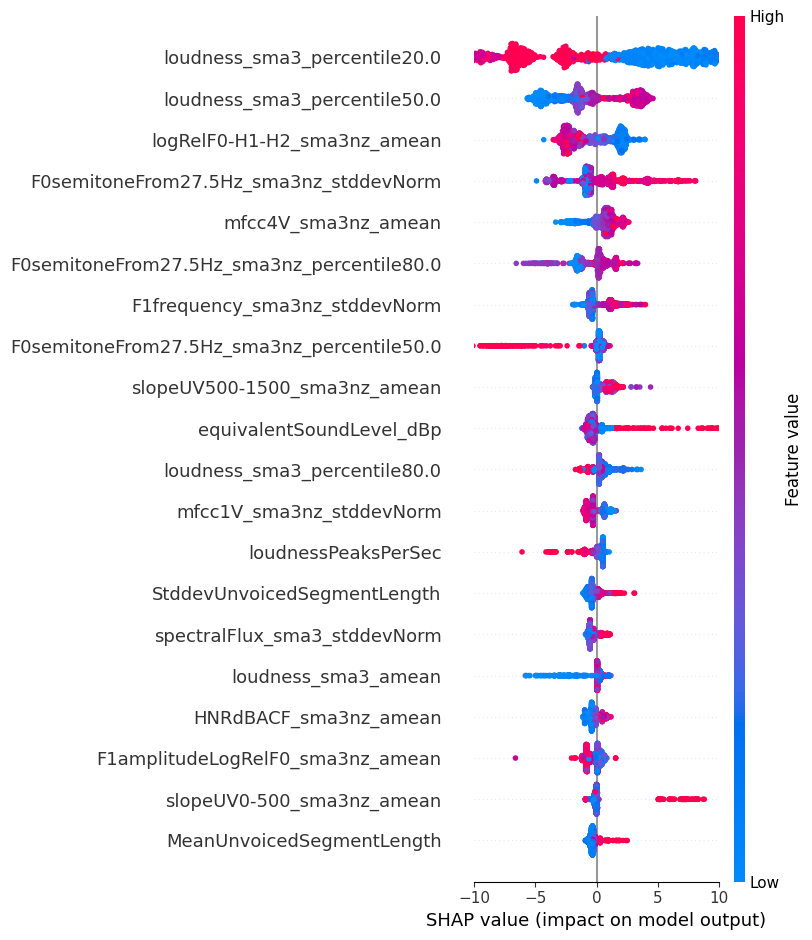}
    \includegraphics[width=0.32\linewidth]{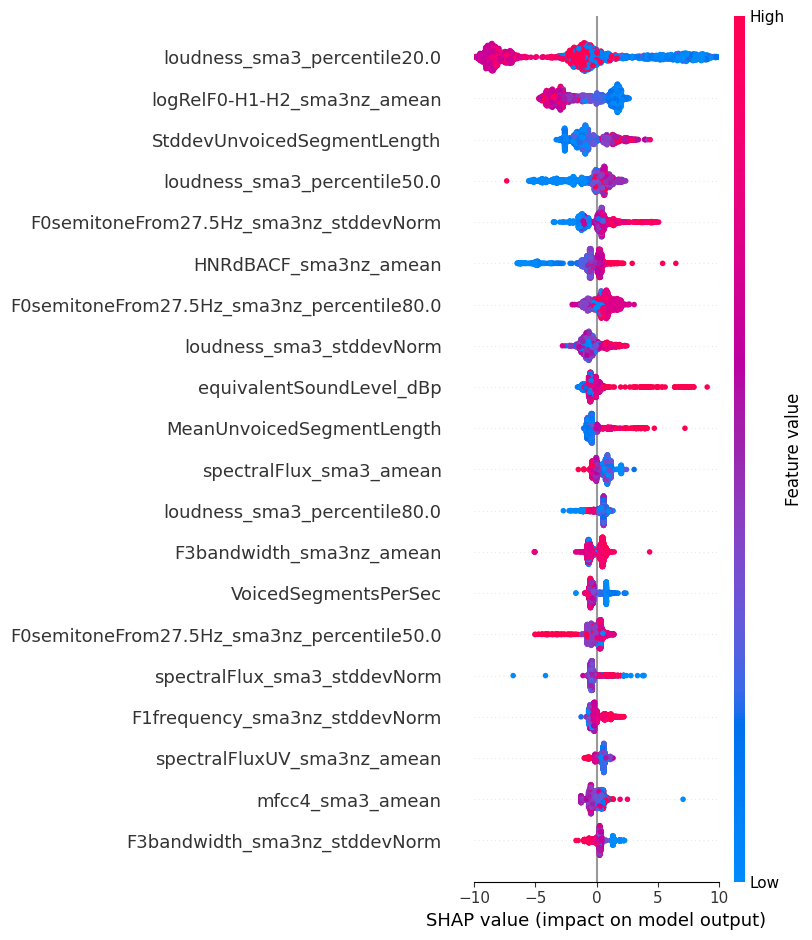}
    \includegraphics[width=0.32\linewidth]{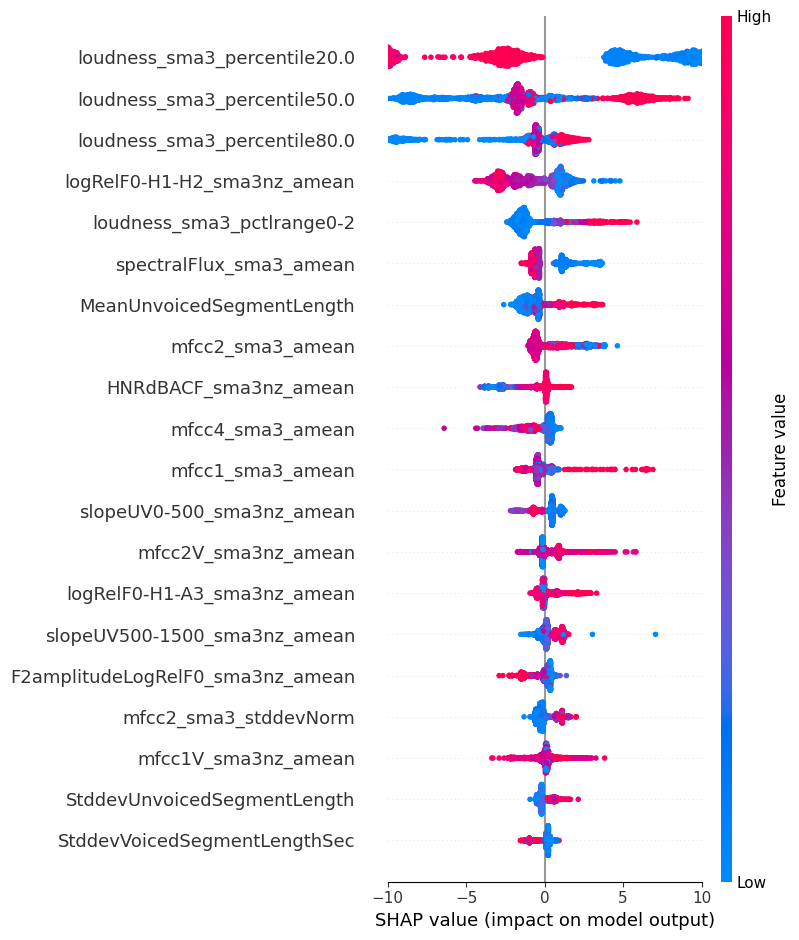}
    \vspace{-.2cm}
    \caption{Shapley values for the TORGO dataset. Left: first second of audio; middle: full audio; right: silent segments}
    \label{fig:shap}
\end{figure*}

\subsection{Explainability}

To further examine model behaviour, we also extracted the Shapley values for the TORGO dataset and the eGeMAPSv02 set (Fig.~\ref{fig:shap}). We can observe that the most important features relate to loudness. Loudness-related features are likely to be influenced by microphone placement, recording gain, speaker-to-microphone distance, and background conditions, making them plausible sources of shortcut learning. To mitigate this, we apply root-mean-square (RMS) normalisation to the TORGO dataset, which yields a F1-score of 0.83 and Shapley values indicate that f1frequency\_sma3nz\_stddevnorm (normalised standard deviation of the first formant on voiced segments) and mfcc3\_sma3\_stddevnorm (normalised standard deviation of the third Mel-Frequency Cepstral Coefficient) are now driving the predictions. .  
However, RMS normalisation does not reduce classification performance for silent segments (F1 = 0.92) or first-second inputs (F1 = 0.83).
Moreover, feature importance profiles vary substantially depending on whether models are trained on full audio, first-second segments, or silence. The absence of consistent feature patterns across conditions indicates that models are unlikely to rely on stable, clinically meaningful acoustic representations. Instead, the importance shifts support the hypothesis that different confounding signals drive classification under different input constraints.

Moreover, removing the first second of audio does not affect performance compared to full audio, indicating that recording bias is present throughout the audio, and sustained speech content alone cannot explain the observed classification performance. This suggests that extracted features may be influenced by recording or environmental factors rather than robust disorder-specific speech characteristics.

To further investigate this effect, we trained models on only the first second of each training recording and evaluated them on full audio from the testing set (Opensmile eGeMAPS). Weighted F1-scores remained comparable to full-audio training (TORGO: 0.73, DAIC: 0.75, UCLASS: 0.84, MDVR-KCL: 0.65, Neurovoz: 0.78), confirming that the first second is enough to generalise to the full audio and that it is unlikely that those features are a useful representation of speech. 

\subsection{Limitations}
This study does not demonstrate that all classification performance arises from confounding factors, nor does it identify the precise sources of the predictive cues. Rather, it shows that substantial predictive information exists outside the intended speech signal across multiple datasets. Future work should investigate the relative contributions of recording artefacts, environmental noise, speaker characteristics, and clinically meaningful speech features.

\section{Conclusion}

Our findings provide evidence that classification performance in several widely used clinical speech datasets may be substantially influenced by non-speech cues and dataset-specific confounding factors.
Across multiple feature representations and input conditions, we found that substantial classification performance could often be achieved using minimal or non-speech segments.
While these results do not imply that disorder-related speech characteristics are uninformative, they demonstrate that non-speech cues can contribute substantially to classification performance and should be carefully considered when developing and evaluating speech-based biomarkers. The consistency of these observations across multiple datasets and disorders suggests that shortcut learning may represent a broader challenge in clinical speech research rather than an isolated issue affecting a single corpus.


Moreover, these findings highlight the importance of detailed methodological reporting (recording device, site, conditions, etc.) and the need for more critical evaluation of widely used datasets.
We recommend that future studies routinely assess performance under constrained conditions, such as short excerpts or non-speech segments, to identify potential confounding factors. Such evaluations may help ensure that reported performance reflects clinically meaningful speech characteristics and improve the reliability and generalisability of speech-based healthcare applications.


\bibliographystyle{IEEEtran}
\bibliography{stuttering}

@article{gomez2025discriminating,
  title={DISCRIMINATING PARKINSON’S DISEASE FROM NORMOPHONIC VOICES USING A SINCNET MODEL},
  author={G{\'o}mez-Garc{\i}a, JA and Fandino-Toro, H and Torricelli, D},
  year={2025}
}

@article{halawani2025enhanced,
  title={Enhanced early skin cancer detection through fusion of vision transformer and CNN features using hybrid attention of EViT-Dens169},
  author={Halawani, Hanan T and Senan, Ebrahim Mohammed and Asiri, Yousef and Abunadi, Ibrahim and Mashraqi, Aisha M and Alshari, Eman A},
  journal={Scientific Reports},
  volume={15},
  number={1},
  pages={34776},
  year={2025},
  publisher={Nature Publishing Group UK London}
}

@article{wu2022classification,
  title={Classification of Alzheimer’s disease based on weakly supervised learning and attention mechanism},
  author={Wu, Xiaosheng and Gao, Shuangshuang and Sun, Junding and Zhang, Yudong and Wang, Shuihua},
  journal={Brain Sciences},
  volume={12},
  number={12},
  pages={1601},
  year={2022},
  publisher={MDPI}
}

@article{di2024machine,
  title={Machine learning-assisted speech analysis for early detection of {P}arkinson’s disease: a study on speaker diarization and classification techniques},
  author={Di Cesare, Michele Giuseppe and Perpetuini, David and Cardone, Daniela and Merla, Arcangelo},
  journal={Sensors},
  volume={24},
  number={5},
  pages={1499},
  year={2024},
  publisher={MDPI}
}

@article{ravi2024enhancing,
  title={Enhancing accuracy and privacy in speech-based depression detection through speaker disentanglement},
  author={Ravi, Vijay and Wang, Jinhan and Flint, Jonathan and Alwan, Abeer},
  journal={Computer speech \& language},
  volume={86},
  pages={101605},
  year={2024},
  publisher={Elsevier}
}

@article{remya2025hybrid,
  title={A Hybrid Cross-Attentive CNN-BiLSTM-Transformer Network for Dysarthria Severity Classification},
  author={Remya, MS and Ishwar, Prakash and Nedungadi, Prema},
  journal={Scientific Reports},
  volume={15},
  number={1},
  pages={42080},
  year={2025},
  publisher={Nature Publishing Group UK London}
}

@article{geirhos2020shortcut,
  title={Shortcut learning in deep neural networks},
  author={Geirhos, Robert and Jacobsen, J{\"o}rn-Henrik and Michaelis, Claudio and Zemel, Richard and Brendel, Wieland and Bethge, Matthias and Wichmann, Felix A},
  journal={Nature Machine Intelligence},
  volume={2},
  number={11},
  pages={665--673},
  year={2020},
  publisher={Nature Publishing Group UK London}
}

@article{haynes2024generalisation,
  title={Generalisation challenges in deep learning models for medical imagery: insights from external validation of covid-19 classifiers},
  author={Haynes, Sophie Crawford and Johnston, Pamela and Elyan, Eyad},
  journal={Multimedia Tools and Applications},
  volume={83},
  number={31},
  pages={76753--76772},
  year={2024},
  publisher={Springer}
}

@article{liu2024clever,
  title={Clever hans effect found in automatic detection of alzheimer's disease through speech},
  author={Liu, Yin-Long and Feng, Rui and Yuan, Jia-Hong and Ling, Zhen-Hua},
  journal={arXiv preprint arXiv:2406.07410},
  year={2024}
}

@article{lombardo2025speech,
  title={Speech analysis and speech emotion recognition in mental disease: a scoping review},
  author={Lombardo, Clara and Esposito, Giulia and Carbone, Silvia and Serrano, Salvatore and Mento, Carmela},
  journal={Frontiers in Psychology},
  volume={16},
  pages={1645860},
  year={2025},
  publisher={Frontiers Media SA}
}

@article{xue2019acoustic,
  title={Acoustic correlates of speech intelligibility. The usability of the eGeMAPS feature set for atypical speech},
  author={Xue, Wei and Cucchiarini, Catia and van Hout, RWNM and Strik, Helmer},
  year={2019}
}

@inproceedings{eyben2010opensmile,
  title={Opensmile: the munich versatile and fast open-source audio feature extractor},
  author={Eyben, Florian and W{\"o}llmer, Martin and Schuller, Bj{\"o}rn},
  booktitle={Proceedings of the 18th ACM international conference on Multimedia},
  pages={1459--1462},
  year={2010}
}

@inproceedings{godino2019approaches,
  title={Approaches to evaluate {P}arkinsonian speech using artificial models},
  author={Godino-Llorente, JI and Moro-Vel{\'a}zquez, L and G{\'o}mez-Garc{\'\i}a, JA and Choi, Jeung-Yoon and Dehak, N and Shattuck-Hufnagel, S},
  booktitle={Automatic Assessment of {P}arkinsonian Speech Workshop},
  pages={77--99},
  year={2019},
  organization={Springer}
}

@article{moro2019forced,
  title={A forced gaussians based methodology for the differential evaluation of {P}arkinson's Disease by means of speech processing},
  author={Moro-Velazquez, Laureano and Gomez-Garcia, Jorge Andres and Godino-Llorente, Juan Ignacio and Villalba, Jes{\'u}s and Rusz, Jan and Shattuck-Hufnagel, Stephanie and Dehak, Najim},
  journal={Biomedical Signal Processing and Control},
  volume={48},
  pages={205--220},
  year={2019},
  publisher={Elsevier}
}

@article{moro2019analysis,
  title={Analysis of phonatory features for the automatic detection of {P}arkinson’s disease in two different corpora},
  author={Moro-Vel{\'a}zquez, Laureano and G{\'o}mez-Garc{\'\i}a, Jorge A and Dehak, Najim and Godino-Llorente, Juan I},
  journal={Proceedings of the Models and Analysis of Vocal Emissions for Biomedical Applications (MAVEBA)},
  pages={33},
  year={2019}
}

@inproceedings{moro2018study,
  title={Study of the automatic detection of parkison’s disease based on speaker recognition technologies and allophonic distillation},
  author={Moro-Velazquez, Laureano and Gomez-Garcia, Jorge Andres and Godino-Llorente, Juan Ignacio and Rusz, Jan and Skodda, Sabine and Grandas, Francisco and Velazquez, Jos{\'e}-Miguel and Orozco-Arroyave, Juan Rafael and Noth, E and Dehak, Najim},
  booktitle={2018 40th Annual International Conference of the IEEE Engineering in Medicine and Biology Society (EMBC)},
  pages={1404--1407},
  year={2018},
  organization={IEEE}
}

@inproceedings{moro2017use,
  title={Use of acoustic landmarks and gmm-ubm blend in the automatic detection of {P}arkinson’s disease},
  author={Moro-Velazquez, L and Godino-Llorente, JI and G{\'o}mez-Garc{\'\i}a, JA and Villalba, J and Shattuck-Hufnagel, S and Dehak, N},
  booktitle={Models and Analysis of Vocal Emissions for Biomedical Applications: 10th International Workshop},
  volume={117},
  pages={73},
  year={2017}
}

@article{moro2019phonetic,
  title={Phonetic relevance and phonemic grouping of speech in the automatic detection of {P}arkinson’s Disease},
  author={Moro-Velazquez, Laureano and Gomez-Garcia, Jorge A and Godino-Llorente, Juan I and Grandas-Perez, Francisco and Shattuck-Hufnagel, Stefanie and Yag{\"u}e-Jimenez, Virginia and Dehak, Najim},
  journal={Scientific reports},
  volume={9},
  number={1},
  pages={19066},
  year={2019},
  publisher={Nature Publishing Group UK London}
}

@article{schalling2018speech,
  title={Speech and communication changes reported by people with {P}arkinson’s disease},
  author={Schalling, Ellika and Johansson, Kerstin and Hartelius, Lena},
  journal={Folia Phoniatrica et Logopaedica},
  volume={69},
  number={3},
  pages={131--141},
  year={2018},
  publisher={S. Karger AG Basel, Switzerland}
}

@article{bloem2021parkinson,
  title={{P}arkinson's disease},
  author={Bloem, Bastiaan R and Okun, Michael S and Klein, Christine},
  journal={The Lancet},
  volume={397},
  number={10291},
  pages={2284--2303},
  year={2021},
  publisher={Elsevier}
}

@article{hassan2025enhanced,
  title={Enhanced dysarthria detection in cerebral palsy and ALS patients using WaveNet and CNN-BiLSTM models: A comparative study with model interpretability},
  author={Hassan, Esraa and Saber, Abeer and Abd El-Hafeez, Tarek and Medhat, T and Shams, Mahmoud Y},
  journal={Biomedical Signal Processing and Control},
  volume={110},
  pages={108128},
  year={2025},
  publisher={Elsevier}
}

@article{sajiha2024automatic,
  title={Automatic dysarthria detection and severity level assessment using CWT-layered CNN model},
  author={Sajiha, Shaik and Radha, Kodali and Venkata Rao, Dhulipalla and Sneha, Nammi and Gunnam, Suryanarayana and Bavirisetti, Durga Prasad},
  journal={EURASIP Journal on Audio, Speech, and Music Processing},
  volume={2024},
  number={1},
  pages={33},
  year={2024},
  publisher={Springer}
}

@article{polur2005experiments,
  title={Experiments with fast Fourier transform, linear predictive and cepstral coefficients in dysarthric speech recognition algorithms using hidden Markov model},
  author={Polur, Prasad D and Miller, Gerald E},
  journal={IEEE Transactions on Neural Systems and Rehabilitation Engineering},
  volume={13},
  number={4},
  pages={558--561},
  year={2005},
  publisher={IEEE}
}

@inproceedings{jothieswari2024enhancing,
  title={Enhancing dysarthria detection: Harnessing ensemble models and MFCC},
  author={Jothieswari, J and Manicka Sundara Valli, T and Suguna, S},
  booktitle={International Conference on Smart Computing and Communication},
  pages={135--147},
  year={2024},
  organization={Springer}
}

@article{muhammad2017voice,
  title={Voice pathology detection using interlaced derivative pattern on glottal source excitation},
  author={Muhammad, Ghulam and Alsulaiman, Mansour and Ali, Zulfiqar and Mesallam, Tamer A and Farahat, Mohamed and Malki, Khalid H and Al-Nasheri, Ahmed and Bencherif, Mohamed A},
  journal={Biomedical signal processing and control},
  volume={31},
  pages={156--164},
  year={2017},
  publisher={Elsevier}
}

@inproceedings{vashkevich2018features,
  title={Features extraction for the automatic detection of ALS disease from acoustic speech signals},
  author={Vashkevich, Maxim and Azarov, Elias and Petrovsky, Alexander and Rushkevich, Yuliya},
  booktitle={2018 Signal Processing: Algorithms, Architectures, Arrangements, and Applications (SPA)},
  pages={321--326},
  year={2018},
  organization={IEEE}
}

@article{brady2011impact,
  title={The impact of stroke-related dysarthria on social participation and implications for rehabilitation},
  author={Brady, Marian C and Clark, Alexander M and Dickson, Sylvia and Paton, Gillian and Barbour, Rosaline S},
  journal={Disability and rehabilitation},
  volume={33},
  number={3},
  pages={178--186},
  year={2011},
  publisher={Taylor \& Francis}
}

@article{darley1969differential,
  title={Differential diagnostic patterns of dysarthria},
  author={Darley, Frederic L and Aronson, Arnold E and Brown, Joe R},
  journal={Journal of speech and hearing research},
  volume={12},
  number={2},
  pages={246--269},
  year={1969},
  publisher={American Speech-Language-Hearing Association}
}

@article{enderby2013disorders,
  title={Disorders of communication: dysarthria},
  author={Enderby, Pam},
  journal={Handbook of clinical neurology},
  volume={110},
  pages={273--281},
  year={2013},
  publisher={Elsevier}
}

@article{esposito2020biological,
  title={The biological face of melancholia: Are there any reliable biomarkers for this depression subtype?},
  author={Esposito, Cecilia Maria and Buoli, Massimiliano},
  journal={Journal of Affective Disorders},
  volume={266},
  pages={802--809},
  year={2020},
  publisher={Elsevier}
}

@article{smirnova2018language,
  title={Language patterns discriminate mild depression from normal sadness and euthymic state},
  author={Smirnova, Daria and Cumming, Paul and Sloeva, Elena and Kuvshinova, Natalia and Romanov, Dmitry and Nosachev, Gennadii},
  journal={Frontiers in psychiatry},
  volume={9},
  pages={105},
  year={2018},
  publisher={Frontiers Media SA}
}

@article{li2023using,
  title={Using deeply time-series semantics to assess depressive symptoms based on clinical interview speech},
  author={Li, Nanxi and Feng, Lei and Hu, Jiaxue and Jiang, Lei and Wang, Jing and Han, Jiali and Gan, Lu and He, Zhiyang and Wang, Gang},
  journal={Frontiers in Psychiatry},
  volume={14},
  pages={1104190},
  year={2023},
  publisher={Frontiers Media SA}
}

@inproceedings{ilias2024cross,
  title={A Cross-Attention Layer coupled with Multimodal Fusion Methods for Recognizing Depression from Spontaneous Speech.},
  author={Ilias, Loukas and Askounis, Dimitris},
  booktitle={Interspeech},
  volume={2024},
  pages={912--916},
  year={2024}
}

@inproceedings{rohanian2019detecting,
  title={Detecting depression with word-level multimodal fusion},
  author={Rohanian, Morteza and Hough, Julian and Purver, Matthew},
  booktitle={Proc. Interspeech 2019},
  pages={1443--1447},
  year={2019}
}

@book{american2015american,
  title={The American Psychiatric Association practice guidelines for the psychiatric evaluation of adults},
  author={American Psychiatric Association and others},
  year={2015},
  publisher={American Psychiatric Association}
}

@article{lepine2011increasing,
  title={The increasing burden of depression},
  author={L{\'e}pine, Jean-Pierre and Briley, Mike},
  journal={Neuropsychiatric disease and treatment},
  volume={7},
  number={sup1},
  pages={3--7},
  year={2011},
  publisher={Taylor \& Francis}
}

@inproceedings{gratch2014distress,
  title={The distress analysis interview corpus of human and computer interviews.},
  author={Gratch, Jonathan and Artstein, Ron and Lucas, Gale M and Stratou, Giota and Scherer, Stefan and Nazarian, Angela and Wood, Rachel and Boberg, Jill and DeVault, David and Marsella, Stacy and others},
  booktitle={LREC},
  volume={14},
  pages={3123--3128},
  year={2014},
  organization={Reykjavik}
}

@article{howell2009university,
  title={The university college london archive of stuttered speech (uclass)},
  author={Howell, Peter and Davis, Stephen and Bartrip, Jon},
  journal={Journal of speech, language, and hearing research},
  volume={52},
  number={2},
  pages={556--569},
  year={2009}
}

@article{cummins2015review,
  title={A review of depression and suicide risk assessment using speech analysis},
  author={Cummins, Nicholas and Scherer, Stefan and Krajewski, Jarek and Schnieder, Sebastian and Epps, Julien and Quatieri, Thomas F},
  journal={Speech communication},
  volume={71},
  pages={10--49},
  year={2015},
  publisher={Elsevier}
}

@article{robin2020evaluation,
  title={Evaluation of speech-based digital biomarkers: review and recommendations},
  author={Robin, Jessica and Harrison, John E and Kaufman, Liam D and Rudzicz, Frank and Simpson, William and Yancheva, Maria},
  journal={Digital biomarkers},
  volume={4},
  number={3},
  pages={99--108},
  year={2020},
  publisher={S. Karger AG}
}

@article{voleti2019review,
  title={A review of automated speech and language features for assessment of cognitive and thought disorders},
  author={Voleti, Rohit and Liss, Julie M and Berisha, Visar},
  journal={IEEE journal of selected topics in signal processing},
  volume={14},
  number={2},
  pages={282--298},
  year={2019},
  publisher={IEEE}
}

@article{de2020artificial,
  title={Artificial intelligence, speech, and language processing approaches to monitoring Alzheimer’s disease: a systematic review},
  author={De la Fuente Garcia, Sofia and Ritchie, Craig W and Luz, Saturnino},
  journal={Journal of Alzheimer’s Disease},
  volume={78},
  number={4},
  pages={1547--1574},
  year={2020},
  publisher={SAGE Publications Sage UK: London, England}
}

@article{bowden2023systematic,
  title={A systematic review and narrative analysis of digital speech biomarkers in motor neuron disease},
  author={Bowden, Molly and Beswick, Emily and Tam, Johnny and Perry, David and Smith, Alice and Newton, Judy and Chandran, Siddharthan and Watts, Oliver and Pal, Suvankar},
  journal={NPJ digital medicine},
  volume={6},
  number={1},
  pages={228},
  year={2023},
  publisher={Nature Publishing Group UK London}
}

@inproceedings{torralba2011unbiased,
  title={Unbiased look at dataset bias},
  author={Torralba, Antonio and Efros, Alexei A},
  booktitle={CVPR 2011},
  pages={1521--1528},
  year={2011},
  organization={IEEE}
}

@article{goetz2024generalization,
  title={Generalization—a key challenge for responsible AI in patient-facing clinical applications},
  author={Goetz, Lea and Seedat, Nabeel and Vandersluis, Robert and van der Schaar, Mihaela},
  journal={NPJ Digital Medicine},
  volume={7},
  number={1},
  pages={126},
  year={2024},
  publisher={Nature Publishing Group UK London}
}

@article{shorten2019survey,
  title={A survey on image data augmentation for deep learning},
  author={Shorten, Connor and Khoshgoftaar, Taghi M},
  journal={Journal of big data},
  volume={6},
  number={1},
  pages={1--48},
  year={2019},
  publisher={Springer}
}

@inproceedings{zhang2018mitigating,
  title={Mitigating unwanted biases with adversarial learning},
  author={Zhang, Brian Hu and Lemoine, Blake and Mitchell, Margaret},
  booktitle={Proceedings of the 2018 AAAI/ACM Conference on AI, Ethics, and Society},
  pages={335--340},
  year={2018}
}

@article{ribeiro2016model,
  title={Model-agnostic interpretability of machine learning},
  author={Ribeiro, Marco Tulio and Singh, Sameer and Guestrin, Carlos},
  journal={arXiv preprint arXiv:1606.05386},
  year={2016}
}

@article{rudzicz2012torgo,
  title={The TORGO database of acoustic and articulatory speech from speakers with dysarthria},
  author={Rudzicz, Frank and Namasivayam, Aravind Kumar and Wolff, Talya},
  journal={Language resources and evaluation},
  volume={46},
  number={4},
  pages={523--541},
  year={2012},
  publisher={Springer}
}

@article{mendes2024neurovoz,
  title={NeuroVoz: a Castillian Spanish corpus of {P}arkinsonian speech},
  author={Mendes-Laureano, Jana{\'\i}na and G{\'o}mez-Garc{\'\i}a, Jorge A and Guerrero-L{\'o}pez, Alejandro and Luque-Buzo, Elisa and Arias-Londo{\~n}o, Juli{\'a}n D and Grandas-P{\'e}rez, Francisco J and Godino-Llorente, Juan I},
  journal={Scientific Data},
  volume={11},
  number={1},
  pages={1367},
  year={2024},
  publisher={Nature Publishing Group UK London}
}

@article{jaeger2019mobile,
  title={Mobile Device Voice Recordings at King’s College London (MDVR-KCL) from both early and advanced {P}arkinson’s disease patients and healthy controls},
  author={Jaeger, Hagen and Trivedi, Dhaval and Stadtschnitzer, Michael},
  journal={Zenodo},
  year={2019}
}

@inproceedings{javanmardi2023wav2vec,
  title={Wav2vec-based detection and severity level classification of dysarthria from speech},
  author={Javanmardi, Farhad and Tirronen, Saska and Kodali, Manila and Kadiri, Sudarsana Reddy and Alku, Paavo},
  booktitle={Icassp 2023-2023 IEEE international conference on acoustics, speech and signal processing (icassp)},
  pages={1--5},
  year={2023},
  organization={IEEE}
}

@article{cai2024voice,
  title={Voice disorder classification using Wav2vec 2.0 feature extraction},
  author={Cai, Jie and Song, Yuliang and Wu, Jianghao and Chen, Xiong},
  journal={Journal of Voice},
  year={2024},
  publisher={Elsevier}
}

@article{choi2025comparative,
  title={Comparative Evaluation of Acoustic Feature Extraction Tools for Clinical Speech Analysis},
  author={Choi, Anna Seo Gyeong and Richardson, Alexander and Partlan, Ryan and Tang, Sunny and Cho, Sunghye},
  journal={Interspeech},
  year={2025}
}

@inproceedings{schu2023dysarthric,
  title={On using the {UA-Speech} and {TORGO} databases to validate automatic dysarthric speech classification approaches},
  author={Schu, Guilherme and Janbakhshi, Parvaneh and Kodrasi, Ina},
  booktitle={ICASSP 2023-2023 IEEE International Conference on Acoustics, Speech and Signal Processing (ICASSP)},
  pages={1--5},
  year={2023},
  organization={IEEE}
}

\end{document}